\documentclass[a4paper]{article}

\usepackage{16lomcon}        
\usepackage{cite}             
\usepackage{graphicx}           

\begin{document}

\title{INCLUSIVE TAU LEPTON HADRONIC DECAY IN \\
THE FRAMEWORK OF DISPERSIVE APPROACH TO QCD}

\author{A.V.~Nesterenko \email{nesterav@theor.jinr.ru}}

\affiliation{BLTPh JINR, Dubna, 141980, Russian Federation}

\date{3 January 2014}
\maketitle

\begin{abstract}
The dispersive approach to QCD and its applications to inclusive
$\tau$~lepton hadronic decay and hadronic vacuum polarization function
are briefly discussed.
\end{abstract}

The kinematic restrictions on a number of the strong interaction processes
are embodied within dispersion relations for such quantities as hadronic
vacuum polarization function~$\Pi(q^2)$, related $R$--ratio of
electron--positron annihilation into hadrons $R(s) = {\rm
Im}\,\Pi(s+i0_{+})/\pi$, and Adler function~\cite{Adler}
$D(Q^2)=-d\,\Pi(-Q^2)/d\ln Q^2$. In turn, dispersion relations impose
intrinsically nonperturbative constraints on the functions on hand, that
should certainly be taken into account when one oversteps the limits of
applicability of perturbation theory. These constraints were properly
accounted for within dispersive approach to QCD~\cite{DQCD1a, PRD88,
DQCD34C12} (its preliminary formulation was discussed in
Ref.~\cite{DQCDPrelim12}), which provides the following integral
representations for the aforementioned functions:
\begin{eqnarray}
\label{P_DQCD}
&&\Delta\Pi(q^2,\, q_0^2) = \Delta\Pi^{(0)}(q^2,\, q_0^2)
+ \!\int_{m^2}^{\infty} \rho(\sigma)
\ln\Bigl(\frac{\sigma-q^2}{\sigma-q_0^2}
\frac{m^2-q_0^2}{m^2-q^2}\Bigr)\frac{d\,\sigma}{\sigma}, \qquad \\[-0.25mm]
\label{R_DQCD}
&&R(s) = R^{(0)}(s) + \theta(s-m^2) \int_{s}^{\infty}\!
\rho(\sigma) \frac{d\,\sigma}{\sigma}, \\[-0.75mm]
\label{Adler_DQCD}
&&D(Q^2) = D^{(0)}(Q^2) + \frac{Q^2}{Q^2+m^2}
\int_{m^2}^{\infty} \rho(\sigma)
\frac{\sigma-m^2}{\sigma+Q^2} \frac{d\,\sigma}{\sigma},
\end{eqnarray}
with~$\rho(\sigma)$ being the spectral density
\begin{equation}
\label{RhoGen2}
\rho(\sigma) = \frac{1}{\pi} \frac{d}{d\,\ln\sigma}\,
\mbox{Im}\, p(\sigma-i0_{+})
= - \frac{d}{d\,\ln\sigma}\, r(\sigma)
= \frac{1}{\pi}\, \mbox{Im}\, d(-\sigma-i0_{+}).
\end{equation}
Here $\Delta\Pi(q^2\!,\, q_0^2) = \Pi(q^2) - \Pi(q_0^2)$, $m$~stands for
the value of the hadronic production threshold, $Q^2 \! = \! -q^2 \!>\! 0$
and $s \!=\! q^2 \!>\! 0$ denote spacelike and timelike kinematic
variables, whereas $p(q^2)$, $r(s)$, and~$d(Q^2)$ are the strong
corrections to~$\Pi(q^2)$, $R(s)$, and~$D(Q^2)$. The leading--order terms
in Eqs.~(\ref{P_DQCD})--(\ref{Adler_DQCD}) read~\cite{QEDABFeynman}:
\begin{eqnarray}
&&\Delta\Pi^{(0)}(q^2,\, q_0^2) =
2\,\frac{\varphi - \tan\varphi}{\tan^3\!\varphi}
- 2\,\frac{\varphi_{0} - \tan\varphi_{0}}{\tan^3\!\varphi_{0}}, \qquad \\[-1mm]
&&R^{(0)}(s) =
\theta(s - m^2)\Bigl(1-\frac{m^2}{s}\Bigr)^{\!\!3/2}, \\[-1mm]
&&D^{(0)}(Q^2) =
1 + \frac{3}{\xi}\Bigl[1 \!-\! \sqrt{1\!+\!\xi^{-1}}\,
\sinh^{-1}\!\bigl(\xi^{1/2}\bigr)\Bigr],
\end{eqnarray}
where $\sin^2\!\varphi = q^2/m^2$, $\sin^2\!\varphi_{0} = q^{2}_{0}/m^2$,
and $\xi=Q^2/m^2$, see also Refs.~\cite{PRD88, DQCD34C12}.

Note that in the massless limit ($m=0$) for the case of perturbative
spectral density [$\rho(\sigma) = \mbox{Im}\; d_{\mbox{\scriptsize
pert}}(-\sigma - i\,0_{+})/\pi$] Eqs.~(\ref{R_DQCD})
and~(\ref{Adler_DQCD}) become identical to those of the so--called
Analytic Perturbation Theory (APT)~\cite{APT} (see also Refs.~\cite{APT1,
APT2, APT3, APT4, APT5, APT6}). However, it is essential to keep the value
of the hadronic production threshold nonvanishing ($m \neq 0$), see
paper~\cite{PRD88} and references therein for the details. In particular,
the distinction between the representation~(\ref{Adler_DQCD}) and its
massless limit was elucidated in Sect.~4 of Ref.~\cite{DQCD1a} and Sect.~3
of Ref.~\cite{DQCD1b}.

So far, there is no method to restore the unique complete expression for
the spectral density~(\ref{RhoGen2}) appearing in
Eqs.~(\ref{P_DQCD})--(\ref{Adler_DQCD}) (discussion of this issue may be
found in, e.g., Refs.~\cite{PRD62PRD64Review, DQCD2}). In this paper the
model~\cite{PRD88, DQCD34C12} for the spectral density is employed:
\begin{equation}
\label{RhoDef}
\rho(\sigma) = \frac{4}{\beta_{0}}\frac{1}{\ln^{2}(\sigma/\Lambda^2)+\pi^2} +
\frac{\Lambda^2}{\sigma}.
\end{equation}
Here $\beta_{0}=11-2n_{\mbox{\scriptsize f}}/3$, $\Lambda$~denotes the QCD
scale parameter, and $n_{\mbox{\scriptsize f}}$~stands for the number of
active flavors. The first term in the right--hand side of
Eq.~(\ref{RhoDef}) is the one--loop perturbative contribution, whereas the
second term represents intrinsically nonperturbative part of the spectral
density (see Refs.~\cite{PRD88, DQCD34C12} for the details).

\begin{table}[t]
\caption{Values of the QCD scale parameter~$\Lambda$~[MeV] obtained
within perturbative and dispersive approaches from recently updated
ALEPH~\cite{ALEPH050608} and OPAL~\cite{OPAL9912} experimental data
on inclusive $\tau$~lepton hadronic decay (one--loop level,
$n_{\mbox{\tiny f}}=3$ active flavors), see Ref.~\cite{PRD88}.}
\label{Tab:RTau}
\vskip2.5mm
\centerline{\begin{tabular}{|p{25mm}|cc|cc|}
\hline
&
\multicolumn{2}{|c|}{Perturbative approach\rule[-5pt]{0pt}{15.5pt}}
&
\multicolumn{2}{|c|}{Dispersive approach}
\\
& ALEPH~\cite{ALEPH050608}
& OPAL~\cite{OPAL9912}\rule[-5pt]{0pt}{5pt}
& ALEPH~\cite{ALEPH050608}
& OPAL~\cite{OPAL9912}
\\ \hline
\centering
Vector channel\rule{0pt}{12.5pt}
& $434_{-127}^{+117}$
& $445_{-230}^{+201}$
& $408 \pm 30$
& $409 \pm 53$
\\[1mm]
\centering
Axial--vector channel\rule[-5.5pt]{0pt}{5.5pt}
& \multicolumn{2}{|c|}{\raisebox{-2mm}{no solution}}
& \raisebox{-2mm}{$418 \pm 35$}
& \raisebox{-2mm}{$409 \pm 61$}
\\ \hline
\end{tabular}}
\end{table}

The dispersive approach to QCD has been successfully applied to the study
of the inclusive $\tau$~lepton hadronic decay~\cite{PRD88, DQCD34C12}. The
obtained results reveal that the dispersive approach is capable of
describing recently updated ALEPH~\cite{ALEPH050608} and
OPAL~\cite{OPAL9912} experimental data on inclusive $\tau$~lepton hadronic
decay in vector and axial--vector channels. The values of QCD scale
parameter~$\Lambda$ evaluated in both channels appear to be nearly
identical to each other (see Tab.~\ref{Tab:RTau}), that bears witness to
the self--consistency of the developed approach.

It is worthwhile to mention also the papers~\cite{TauAPT12, TauAPT3},
which study the inclusive $\tau$~lepton hadronic decay within massless APT
and its modifications. However, those papers basically deal either with
the total sum of vector and axial--vector terms of the semileptonic
branching ratio or with its vector term only.

\begin{figure}[t]
\centerline{\includegraphics[width=52.5mm,clip]{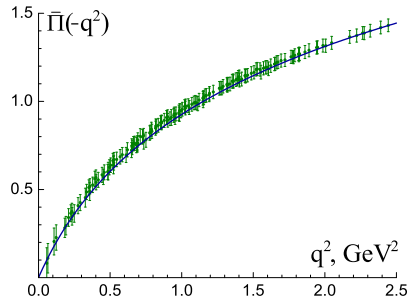}}
\caption{Comparison of the hadronic vacuum polarization
function~(\ref{P_DQCD}) [$\bar{\Pi}(q^2)=\Delta\Pi(0,q^2)$, solid curve]
with relevant lattice simulation data~\cite{Lat2} (circles). The presented
results correspond to the spectral density~(\ref{RhoDef}) and
$n_{\mbox{\tiny f}}=2$ active flavors, see also Refs.~\cite{PRD88, Prep}.}
\label{Plot:PDQCD}
\end{figure}

The dispersive approach to QCD~\cite{DQCD1a, PRD88, DQCD34C12} provides
the representations~(\ref{P_DQCD})--(\ref{Adler_DQCD}), which conform with
the results obtained in Ref.~\cite{PRL99PRD77}. Besides, it was explicitly
shown in Refs.~\cite{DQCD1a, DQCD1b, DQCD2} that the Adler
function~(\ref{Adler_DQCD}) complies with corresponding experimental
prediction in the entire energy range (note that a close matter was
studied in Refs.~\cite{Maxwell, Fischer, Cvetic1, Cvetic2}). Additionally,
as one can infer from Fig.~\ref{Plot:PDQCD}, the hadronic vacuum
polarization function~(\ref{P_DQCD}) is in a good agreement with relevant
low--energy lattice simulation data~\cite{Lat2}, see Refs.~\cite{PRD88,
Prep} for the details.

\end{document}